\pdfoutput=1
\documentclass[amsmath,amsfonts,amssymb,onecolumn,superscriptaddress,11pt]{revtex4}
\usepackage{graphicx,dcolumn,bm,verbatim,hyperref,subfigure}

\begin{document}

\bibliographystyle{apsrev} 

\title{Detection of hidden structures on all scales
in amorphous materials and complex physical systems:
basic notions and applications to networks, lattice systems, and glasses}

\author{P. Ronhovde}
\affiliation{Department of Physics, Washington University in St. Louis, Campus Box 1105, 1 Brookings Drive, St. Louis, MO 63130, USA}
\author{S. Chakrabarty}
\affiliation{Department of Physics, Washington University in St. Louis, Campus Box 1105, 1 Brookings Drive, St. Louis, MO 63130, USA}
\author{M. Sahu}
\affiliation{Department of Physics, Washington University in St. Louis, Campus Box 1105, 1 Brookings Drive, St. Louis, MO 63130, USA}
\author{K. K. Sahu}
\affiliation{Department of Physics, Washington University in St. Louis, Campus Box 1105, 1 Brookings Drive, St. Louis, MO 63130, USA}
\affiliation{Metal Physics and Technology, ETH, 8093 Zurich, Switzerland} 
\author{K. F. Kelton}
\affiliation{Department of Physics, Washington University in St. Louis, Campus Box 1105, 1 Brookings Drive, St. Louis, MO 63130, USA}
\author{N. A. Mauro}
\affiliation{Department of Physics, Washington University in St. Louis, Campus Box 1105, 1 Brookings Drive, St. Louis, MO 63130, USA}
\author{Z. Nussinov$^{*}$}
\affiliation{Department of Physics, Washington University in St. Louis, Campus Box 1105, 1 Brookings Drive, St. Louis, MO 63130, USA}
\affiliation{Kavli Institute for Theoretical Physics, Santa Barbara, CA93106 }
\email[Corresponding author:]{zohar@wuphys.wustl.edu} 



\begin{abstract}
Recent decades have seen the discovery of numerous complex materials.  At the root of the complexity underlying many of these materials lies a large number of possible contending atomic- and larger-scale configurations and the intricate correlations between their constituents.
For a detailed understanding, there is a need for tools that enable the detection of pertinent structures on all spatial and temporal scales. Towards this end, we suggest a new 
method by invoking ideas from network analysis and information
theory. Our method efficiently identifies basic unit cells and topological defects in lattices with low disorder and may analyze general amorphous systems to identify candidate natural structures where a clear definition of order is lacking. This general unbiased detection of physical structure does not require a guess as to which of the system properties should be deemed as important and may constitute a natural point of departure for further analysis. The method applies to both static and dynamic systems.
\end{abstract}
\maketitle

\section{Introduction}
Currently, there are no universal tools for examining complex physical systems in a general and systematic way that fleshes out their pertinent features from the smallest fundamental unit to the largest scale encompassing the entire system.  The challenge posed by these complex materials is acute and stands in stark contrast to that in simple ordered systems. In crystals,  atomic unit cells replicate to span the entire system. Historically, the regular shapes of some large-scale single crystals were suggested to reflect the existence of an underlying repetitive atomic scale unit cell structure long before modern microscopy and the advent of scattering and tunneling techniques. This simplicity enables an understanding of many solids in great detail. In complex systems, rich new structures may appear on additional intermediate scales. Currently, some of the oldest and, after several millennia, heavily investigated complex materials are glasses. Much more recent challenges include the high temperature cuprate and pnictide superconductors, heavy fermion compounds, and many other compounds including, e.g., the manganites, the vanadates, and the ruthenates. These systems exhibit a rich array of behavior including superconductivity and metal to insulator transitions, rich magnetic characteristic and incommensurate orders, colossal magneto-resistance, orbital orders, and novel transport properties. 

A wealth of experimental and numerical data has been accumulated on such systems. The discovery of the salient features in complex materials such as these and more generally of
complex large scale physical systems across all spatial resolutions may afford clues for a more accurate understanding.  In disparate arenas, important guesswork needs to be invoked as to which of the many features of the physical systems are important and may form the foundation for a detailed analysis. With ever-increasing experimental and computational data, such challenges will only sharpen in the coming years. There is a need for methods that may pinpoint central features on all scales. This works suggests a path towards the solution of this problem in complex amorphous materials. A companion work \cite{ronhovde}
provides many of the details not provided in this brief summary. An explanation of our core idea 
require a few concepts from the physics of glasses and network analysis. Towards that end, we review these concepts
below. 

\section{A paradigm for complex systems: glasses}
 \label{sec:glassesintroduction}
 We illustrate the basic premise of our approach by in this work by focusing on glasses.  We begin by, all too briefly, reviewing a central problem-- the detection of natural scales and structures in glasses.  Complex systems such as glasses are not easy to analyze with conventional theoretical tools. As all interactions between the basic constituents of a gas are weak, a gas is easy to understand and analyze. At the other extreme, although the interactions in regular periodic solids are generally strong, such solids may be characterized by their unit cells and related broken symmetries. The situation is, however, radically different for liquids and glasses.
Liquids that are rapidly cooled (``supercooled'') below their melting 
temperature cannot crystallize and instead 
at sufficiently low temperatures become ``frozen'' on experimental times scales 
in an amorphous state (a ``glass''). 
On supercooling, liquids may veer towards local low energy structures \cite{Bernal,Miracle}, 
such as icosahedral structures observed in metallic glasses 
\cite{ref:schenkISRO,ref:keltonfirst}, before being quenched into this
amorphous state.
Because of the lack of a simple crystalline reference, the general structure 
of glasses is notoriously difficult to quantify in a meaningful way beyond the very local 
scales. As such it remains a paradigm for the analysis of structure
in complex materials.  

The most familiar and oldest technological glasses are the common silicate 
glasses.  
More modern glasses include phosphate glasses (biomedical applications), 
semiconductor chalcogenide glasses (optical recording media), 
and metallic glasses (which have high tensile strength and are stronger than steel, resistant to corrosion and wear and tear,  have magnetic properties,  and an extremely high coefficient of restitution). \cite{industry,newscientist,zrwear,moli}  Existing work on glasses is vast. Glass formers display several key features \cite{ref:lubchenkowolynes}. A prominent feature of glass formers is that the viscosity and relaxation times can increase by many orders 
of magnitude over a narrow temperature range. This slowing of dynamics is not accompanied by the thermodynamic signatures of conventional phase transitions
nor a clearly visible pronounced change of spatial structure. 
The high number of metastable energy states in these systems
\cite{ref:angelaniPEL,ref:parisiglasstrans} leads to rich energy 
landscapes\cite{ref:angelaniPEL,ref:parisiglasstrans,ref:sastry,ref:debenedetti,ref:lubchenkoaging,ref:johnsonMRS,ref:doyestructures}. A notable facet of the glass transition reflecting structure in the space-time domain is that of
``dynamical heterogeneity'' \cite{dh1,dh2,dh3,dh4,dh5,dh6,dh7,dh8,dh9,overview}: the fact that the dynamics in supercooled liquids is spatially non-uniform. Many theories of glasses, e.g., \cite{ref:lubchenkowolynes,ktw,tm,mode_coupling,davidr, dyn_con,ref:nussinovAPT,ref:tarjusAPT}
have been advanced over the years.
The theory of random first order transitions (RFOT) 
investigates mosaics of local configurations
\cite{ref:lubchenkowolynes,ktw}.  As shown in \cite{ref:nussinovAPT}, RFOT is
related to theories of ``locally preferred structures'' \cite{ref:nussinovAPT,ref:tarjusAPT,dk,ref:nelsonGF,ref:sadocmosseri}-
which, as befits their name, also rely on the understanding of natural structures in glasses. 
Other theories seek a similar quantification of structure.  Investigations include spin glass approaches \cite{tm}
topological defects and 
kinetic constraints
\cite{ref:nussinovAPT,ref:tarjusAPT,ref:ritortsollich,ref:cvetkovicNZ,ref:aharonov},
and numerous ingenious approaches summarized in excellent reviews, e.g., 
\cite{ref:debenedetti,ref:rev_bert,chandler}.
There is a proof that a growing static length scale must accompany 
the diverging relaxation times of glass \cite{ref:montanariCL}. 
Some evidence has been found for growing correlation lengths (static and those
describing dynamic inhomogeneities)
\cite{tanaka,ref:mosayebiCLSGT,ref:berthierCL,ref:karmakarsastry}.
Correlation lengths were studied via ``point-to-set'' 
correlations \cite{ps} and pattern repetition size \cite{kl}.  Current common methods of characterizing structures that center on an atom or a given link include (a) Voronoi polyhedra, \cite{ref:aharonov,sheng,ref: finney}, (b) Honeycutt-Andersen indices \cite{HA},  and (c) bond orientation \cite{BO}.   A long-standing challenge addressed in this work is the direct detection of structures {\em of general character and scale} in amorphous physical systems. Towards this end, we briefly introduce specific concepts from network analysis. 

\section{Network analysis}

Network analysis has been transformative in generating keen new insights in many areas.  The ideas that we introduce here bring to bear network methods that have been so useful in the social and biological sciences to complex physical problems that have not yet been examined before through this prism. To address the challenge of detecting and characterizing structure in complex systems on all scales, we specifically introduce methods from the new growing physics discipline of  ``community detection'' \cite{phystoday}.  Our key idea is that any complex physical system may be expressed as a network 
of nodes (e.g., atoms, electrons, etc.) and connecting links.
With this representation, we may then apply \emph{multiresolution} methods 
\cite{ref:rzmultires} from network theory to the analysis of complex materials.

\subsection{Partitions of large systems into weakly coupled elements}

\begin{figure}
\centering
\includegraphics[width=.4\columnwidth]{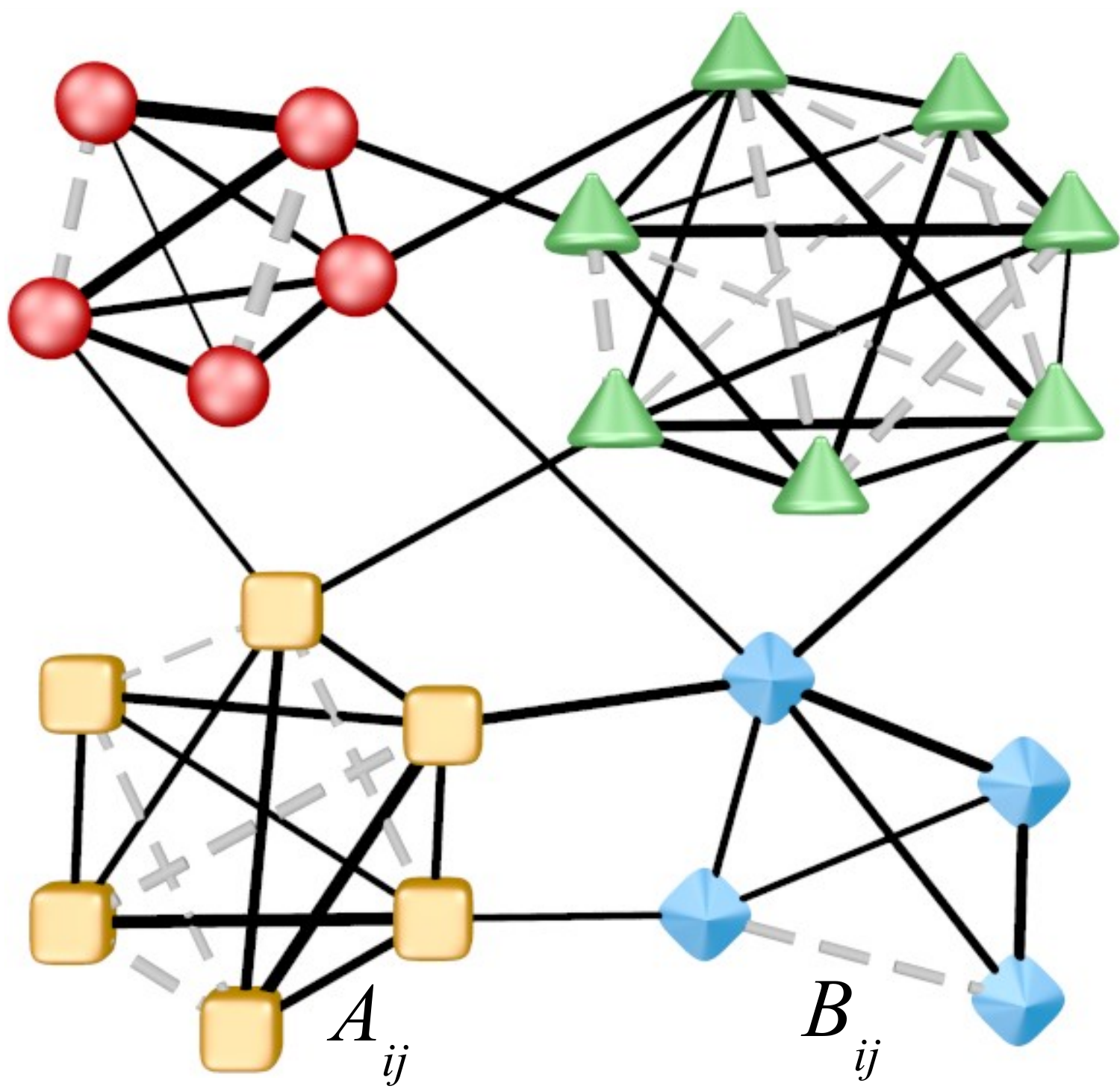}\hspace{0.2cm}%
\caption{From \cite{ronhovde}. A weighted network with $4$ 
natural (strongly connected) communities.
The goal in community detection is to identify such strongly related 
clusters of nodes.
Solid lines depict weighted links corresponding to complimentary or attractive
relationships between nodes $i$ and $j$ (denoted by $A_{ij}$) [$(V_{ij}-v)<0$ 
in Eq.(\ref{eq:ourPottsmodel})].
Gray dashed lines depict missing or repulsive edges
(denoted by $B_{ij}$) [$(V_{ij}-v)>0$].
In both cases, the relative link weight is indicated by the 
respective line thicknesses.}
\label{network1}
\end{figure}

\begin{figure}
\centering
\includegraphics[width=.35\columnwidth]{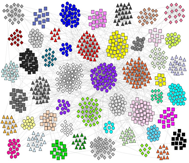}\hspace{0.2cm}%
\includegraphics[width=.38\columnwidth]{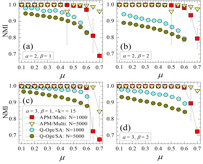}
\caption{From \cite{ref:rzmultires}. A benchmark demonstrating the accuracy of our method. Left: The benchmark generates networks with $N = 1000$ or $5000$ nodes which are assigned to communities of varying sizes specified by a power law distribution parameter $\beta$. The fraction of edges that each node has connected to nodes outside its own community is $\mu$.  The 
coordination numbers of the nodes are given by a power law distribution specified by $\alpha$. Right:  The solutions by our method (``Absolute Potts Model'' (APM))  and those found by simulated annealing (SA) optimizing a popular cost function (``modularity'') are compared to a known answer by  the Normalized Mutual Information (NMI). Perfect agreement corresponds to a value of NMI = 1. }
\label{network}
\end{figure}

{\em Community detection} describes the problem of finding clusters--``communities'' -- groups of nodes 
 with strong internal connections and weak connections between different clusters
 (see Figs.(\ref{network1}, \ref{network})).  
The definitions of nodes and edges depend on the system 
being modeled.
Between each pair
of nodes $i$ and $j$ we have an edge weight $V_{ij}$ which may emulate 
an interaction energy or measured correlation between sites $i$ and $j$ \cite{cov1,cov2,cov3}.
The nodes belong to $q$ communities: $\{C_{a}\}_{a=1}^{q}$.
In our particular physics realization, the nodes represent particles and edges their pair-wise interactions. In an ideal decomposition of
a large graph into completely disjoint communities (groups of particles), there are no interactions between  different communities; the system is effectively that of an ``ideal gas''
of the decoupled communities of particles. In practice,
the task is to find a partition into communities which maximally decouple. 
Such a separation may afford insight into large physical systems. Many approaches to community detection exist, e.g.,
\cite{fortunato1,newman_girvan,blondel,newman_fast,RB,gudkov,RosB,book_comm}  and may be invoked in sociology, homeland security, and other networks \cite{phystoday,real1,real2,real3}. Two of us earlier developed a method \cite{fortunato1} that does not suffer from the  ``resolution limit''  that hinders many approaches \cite{resolution_limit} and that, alongside  \cite{blondel}, 
has been applied to investigate systems with more
than a billion links. Earlier works describing the method and publicly released code, 
\cite{ref:rzmultires,ref:rzlocal,dandan,my_Web}  contained elaborate and precise technical definitions 
to notions briefly reviewed below.

%

\subsection{Our community detection method in a nutshell}

A parameter $\gamma$ in the Hamiltonian below defines the ``resolution'' of the system. \cite{ref:rzmultires,ref:rzlocal,dandan}  
We generalize our earlier works by adding
a background $v$ and allowing for continuous weights $V_{ij}$ instead of discrete weights that are prevalent in graph theory. Our (Potts type) Hamiltonian reads 
\begin{equation}  \label{eq:ourPottsmodel}
	H = \frac{1}{2} \sum_{a=1}^{q} \sum_{i, j \in C_{a}} (V_{ij}-v) [ \theta(v-V_{ij}) 
	+ \gamma \theta(V_{ij}-v)].
	\end{equation}
	In Eq. (\ref{eq:ourPottsmodel}), in the inner sum both nodes $i$ and $j$ belong to the same 
	community $C_{a}$. The outer sum is performed over the $q$ different communities, 
	The number of communities $q$ may be specified from the outset {\em or left arbitrary} (as in
	our multi-resolution method below) and have
	the algorithm decide by steadily increasing the number of communities $q$ for which 
	we have low energy solutions. \cite{ref:rzmultires,ref:rzlocal} 
Minimizing this Hamiltonian corresponds to identifying 
strongly connected clusters of nodes.
The parameter  $\gamma \ge 0$ tunes the relative weights of the connected and unconnected edges and, as advertised above, allows us to vary the targeted scale of the communities 
sought (the system ``resolution''). As seen from Eq.(\ref{eq:ourPottsmodel}), a high value of $\gamma$ leads to forbidding energy 
penalties unless {\em all} intra-community nodes ``attract'' one another [i.e., $(V_{ij} -v)<0$
for all $(ij) \in {\cal{C}}_{a}$], whereas  $\gamma =0$ would not penalize the inclusion of any additional nodes in a given community and the lowest energy solution generally corresponds to the entire physical system 
The model for the current application could be further generalized 
by incorporating $n$-body interactions or correlation functions (such as three or four 
point correlation functions).
Details concerning a greedy minimization of Eq. (\ref{eq:ourPottsmodel}) 
appear in \cite{ref:rzmultires,ref:rzlocal}.
Somewhat better optimization could be obtained with a heat bath
algorithm \cite{dandan} at a cost of a substantially increased computational effort.

\subsection{Multiresolution network analysis} 
\label{mrasec}
We addressed multi-scale partitioning \cite{ref:rzmultires} by employing information-theory measures \cite{info11,info2,info3}  to examine contending partitions for each system scale.  Decreasing $\gamma$, we minimize Eq.(\ref{eq:ourPottsmodel}) with progressively lower intra-community edge densities, effectively ``zooming out'' toward larger structures. A key construct in our approach is that of 
{\em replicas}--  independent solutions of the same problem.
This number of replicas $p$ may be set by the user; a higher value of $p$ leads to
more accurate analysis. 
In earlier work \cite{ref:rzmultires} concerning a static network, replicas were related to one another by permuting the numbers of the nodes that form the very same network. However, many other definitions of replicas
can be considered (as for the analysis of dynamics in complex physical systems).
We can automatically determine all the natural scales of the system by identifying the values of $\gamma$ for which these replicas agree most strongly as seen via measures of information theory overlaps.  
We briefly elaborate on these measures in the current context \cite{ref:rzmultires}.
The probability for a randomly selected node to be in a community $a$ is $P(a) = n_a/N$ with $n_a$  the number of nodes in community $a$ and $N$ the total number of nodes . If there are q communities in a partition A, then the Shannon entropy is  $H_{A} = - \sum_{a=1}^{q}  \frac{n_{a}}{N} \log_{2} \frac{n_{a}}{N}$.
The mutual information $I(A,B)$ between solutions (partitions)  found by two replicas A and B is $I(A,B) = \sum_{a=1}^{q_{A}} \sum_{b=1}^{q_{B}} \frac{n_{ab}}{N} \log_{2} \frac{n_{ab} N}{n_{a} n_{b}}$. Here,
 $q_A$ and $q_B$ are the number of communities in partitions A and B, $n_{ab}$ is the number of nodes of community $a$ of partition A that are shared with community $b$ of partition B, $n_a$ is the number of nodes in community $a$ of partition A, and $n_b$ is the number of nodes in community $b$ of partition B.  The variation of information between two partitions A and B is given by $VI(A,B) = H_{A} + H_{B} - 2I(A,B)$  ($0 \le VI(A,B) \le  \log_{2} N$). The Normalized Mutual Information 
is defined as $NMI(A,B) = \frac{2I(A,B)}{H_{A} + H_{B}}$  ($0 \le NMI(A,B) \le 1$).
A high average NMI indicates high agreement between different replicas.
The VI measures the disparity between different replicas. A low VI indicates high agreement between different replicas. A large VI indicates a high variance- large fluctuations between the results
found in different replicas. The central result of \cite{ref:rzmultires} was that
{\em extrema (including plateaux) of information theory overlaps when averaged over all replica pairs,
(e.g., the average NMI: $ I_{N} = \frac{2}{p(p-1)} \sum_{A \neq B} NMI(A,B)$ and the average
VI) indicate the natural network scales.} \cite{ref:rzmultires} That is, we may find the values
$\gamma^*$ for which the average  $Q$ of information theory overlaps
such as $NMI$ and $VI$ over all replica pairs
 $(dQ/d \gamma)|_{\gamma = \gamma^{*}} =0$ and then 
determine the minima of Eq. \ref{eq:ourPottsmodel} for these $\gamma^*$(s).
The method identifies all ``natural'' scales of the system. This approach is fast
\cite{ref:rzmultires,ref:rzlocal} and has an accuracy that surpasses methods such 
as simulated annealing (SA) applied to disparate cost functions \cite{ref:rzmultires,ref:rzlocal}
(see also Fig. 1). More notably, to our knowledge, this approach is the only one that quantitatively evaluates the ``natural'' partitions over all scales. Other current approaches to (non-multi scale) community detection include optimizing ``modularity'', \cite{newman_girvan} dynamics in high dimensions, \cite{gudkov}  data compression, \cite{RosB}. and numerous other ideas. A detailed analysis Ref. \cite{fortunato1} compared the accuracy of several algorithms for non multi-scale community detection. Multiresolution approaches 
\cite{ref:rzmultires,arenas,fortunato2} are far more recent. By relatively trivial extensions, \cite{ronhovde} the method of \cite{ref:rzmultires} can be applied to the detection of overlapping communities.

\section{Detection of multi-scale structures (static and dynamic) in complex systems}  

\begin{figure}
\centering
\includegraphics[width=.35\columnwidth]{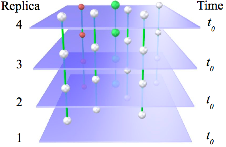}\hspace{0.2cm}%
\includegraphics[width=.38\columnwidth]{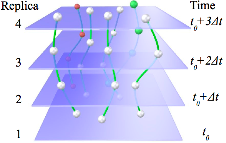}
\caption{From \cite{ronhovde}. Left: our replica construction for the physical system 
at a ``static'' time $t_0$ (with no time separation between replicas).
Right: a similar set of replicas separated by a time $\Delta t$ 
between successive replicas.
We generate a model network for each replica using 
the potential energy between the atoms as the respective edge weights and then solve each replica independently by minimizing Eq. (\ref{eq:ourPottsmodel}) 
at a given value of $\gamma$.
We then use information measures \cite{ref:rzmultires} to evaluate 
how strongly pairs of replicas agree on the ground states of Eq.(\ref{eq:ourPottsmodel}).}
\label{fig:MRAreplicaspic} 
\end{figure}

We wish to analyze complex systems to ascertain general hidden structure in a general
manner with {\em no prior assumptions} as to what the important system
properties may be. To achieve this, we cast physical systems as networks representing atoms (or electrons etc.) as nodes and setting the graph edge weights  in Eq.(\ref{eq:ourPottsmodel}) to be either (i) pair interaction energies or (ii) experimentally measured inter-node (inter-atomic) correlations. As reviewed in Section \ref{mrasec},  our approach to multi-scale community detection  \cite{ref:rzmultires} is simple: 
copies of the community detection problem are given to different ``solvers'' (or ``replicas'').  If the starting points of different replicas in the complex energy landscape are different then they will generally arrive at different solutions (different community groupings). If many of these solvers strongly agree about some features of the solution, then these aspects are more likely to be correct manifesting in extrema of their information theory correlations-\cite{ref:rzmultires}.
When applying this to a physical system, the replicas can be chosen to be copies of the system all at the same time in order to detect natural static scales and structures (panel (a) of 
Fig. \ref{fig:MRAreplicaspic}).  Alternatively, the replicas may be copies of the system at different times as in panel (b) of 
Fig. \ref{fig:MRAreplicaspic} enabling the detection of general spatio-temporal correlations. In both the static and dynamic cases, we find the extrema of the information theory correlations as a function of $\gamma$ the "resolution" parameter in our hamiltonian Eq. (\ref{eq:ourPottsmodel}). Once these extremal values of $\gamma$ are found, the ground states of Eq.(\ref{eq:ourPottsmodel}) 
determine the pertinent structures as in \cite{ref:rzmultires}.  Multiple extrema in the information theory correlations 
suggest \emph{multiple relevant length/time scales}. In this way, our analysis is not limited to the assumption of one or
two specific correlation lengths relative to which scaling type analysis may be done or what correlation function
should be constructed, etc. Rather, viable natural scales of the system appear as extrema in the calculation of the direct 
information theory overlaps. 

\section{Benchmarks:  crystals, crystals with defects, and spin systems}
\begin{figure}
\centering
\includegraphics[width=.4\columnwidth]{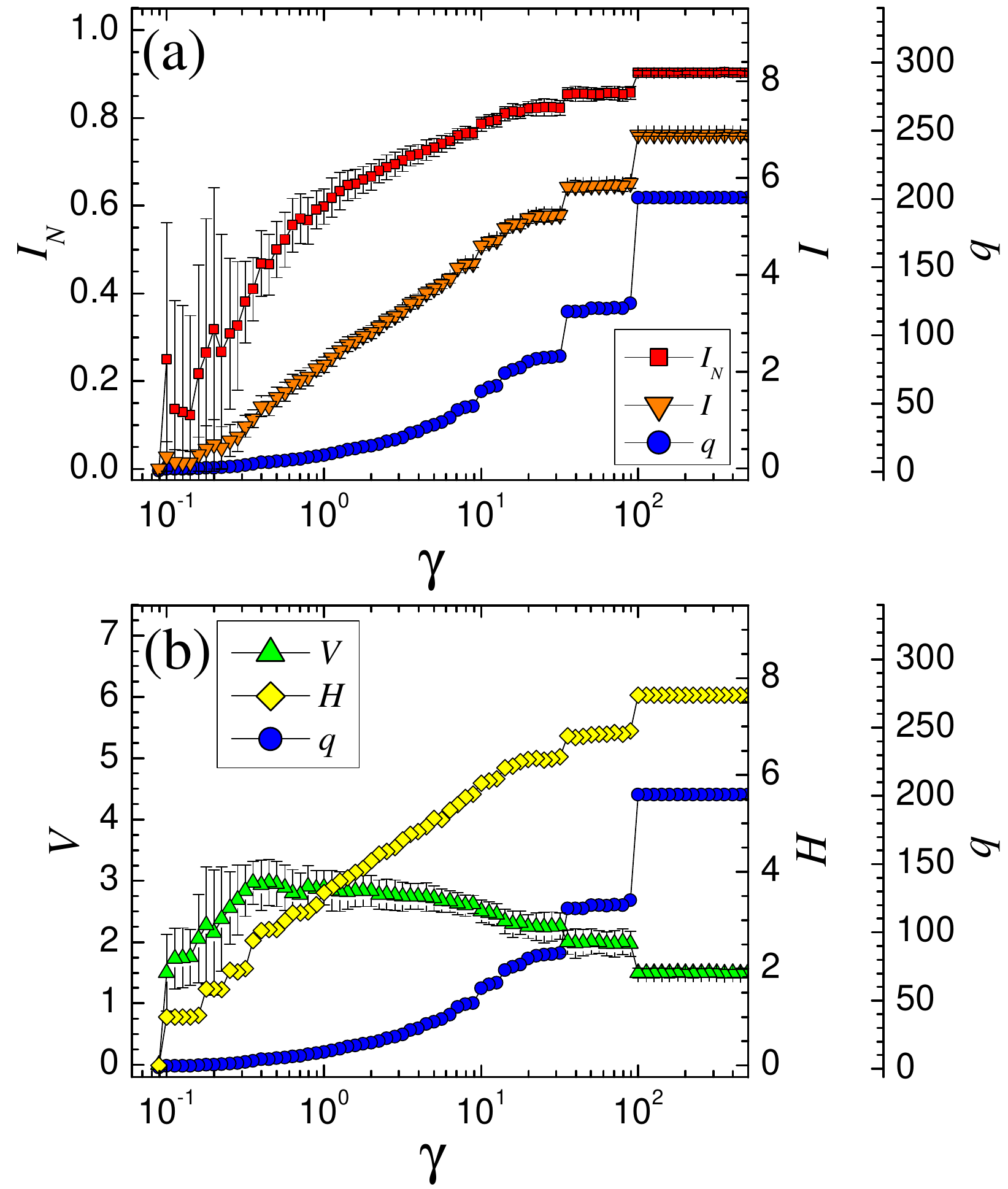}
\includegraphics[width=.4\columnwidth]{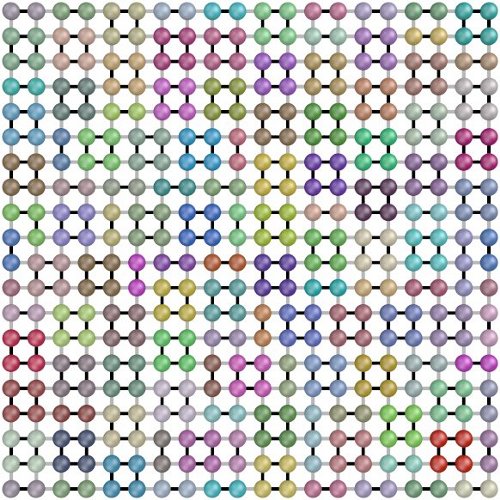}
\caption{From \cite{ronhovde}.
Multiresolution analysis 
of a square lattice with periodic boundary conditions treated as an unweighted graphs as in earlier  analysis \cite{ref:rzmultires}:
Neighbors have an initial weight of $(V_{ij}-v)=1$ in Eq. \ref{eq:ourPottsmodel} and non-neighbors 
have an initial weight $0$. Left: The panels show the information theoretic overlaps between the different replicas when averaged over all replica pairs (see text). These are the variation of information (VI), mutual information (I), normalized mutual information ($I_{N}$), entropy (H) and number of clusters (q) in individual partitions. 
Right:  
Corresponding partition of the
lattice. We use the algorithm described at $\gamma = 60$.
In this configuration, there were $q=120$ clusters with $78$ 
squares, $4$ triads, and $38$ dyads which indicates that square 
configuration dominates the partition, and it shows how our algorithm
can naturally identify the basic unit cells of the square lattice.}
\label{square}
\end{figure}

\begin{figure}
\centering
\includegraphics[width=.4\columnwidth]{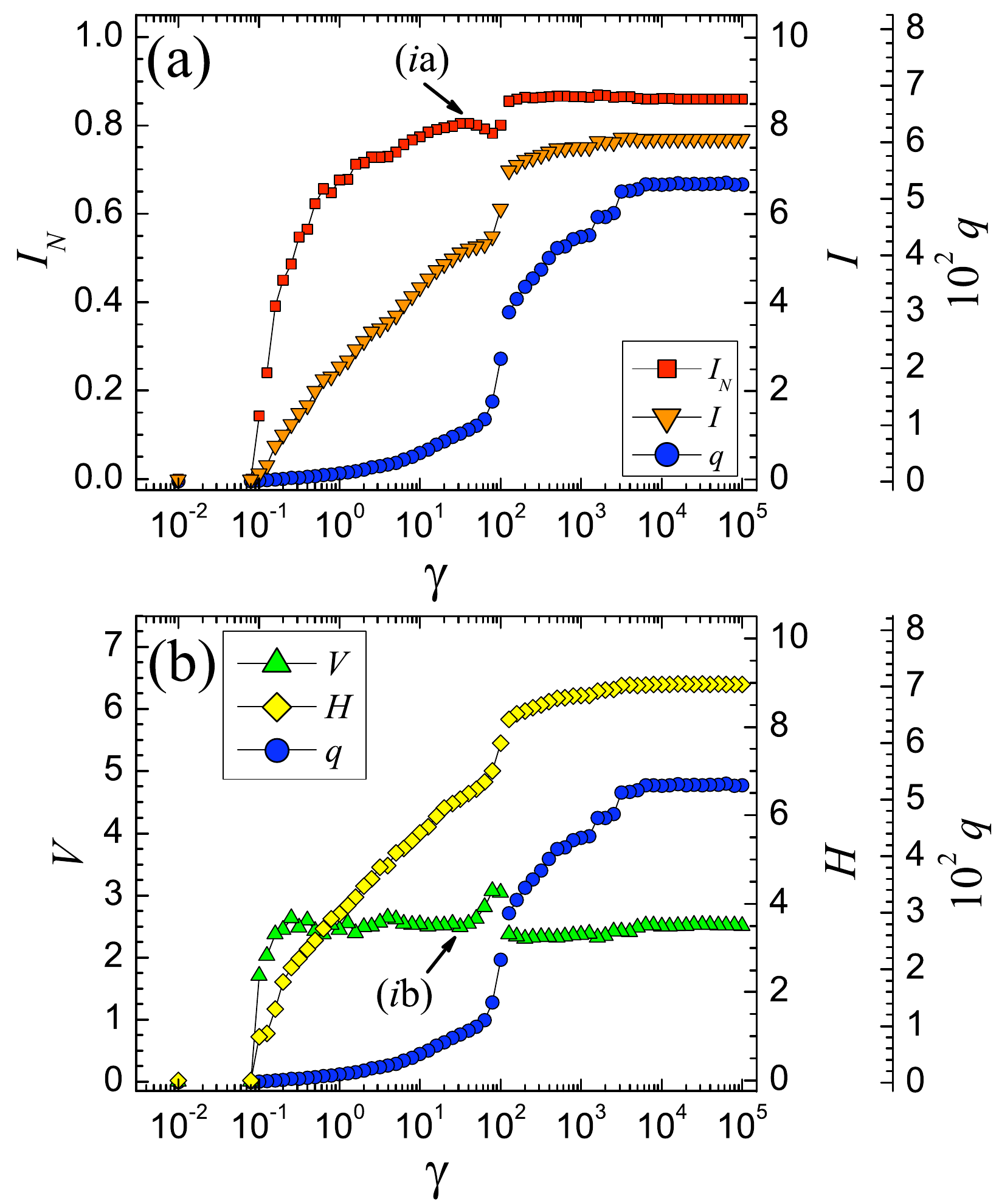}
\includegraphics[width=.4\columnwidth]{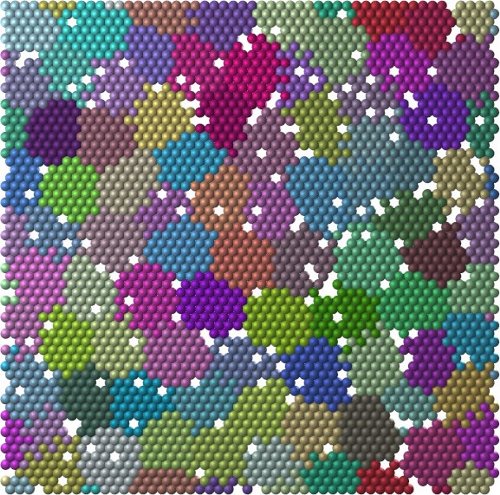}
\caption{From \cite{ronhovde}. Left: A plot for the multiresolution analysis 
of a 2D triangular LJ lattice with periodic boundary conditions.
The legend for the information theory quantities is as in earlier plots.
Edges are weighted according the LJ potential.
there are two preferred regions, a small peak on the left and a large plateau 
on the right, where the peak here corresponds to the largest possible 
``natural'' clusters.
Right:
We use the algorithm at $\gamma\simeq 31.6$
(the left peak) to solve the system.
Our method generally {\em places defects near the boundaries}
of the communities in order
to minimize the energy cost.}
\label{defects}
\end{figure}

 Before applying our method to complex systems, we need to make sure that it yields sensible results in simple physical cases
 aside from the standard networks such as those of Fig. \ref{network}.  
 Hence, we tested it for: 
{\bf (i)} Lattices viewed as graphs, {\bf (ii)} Lattices with
defects, and  {\bf (iii}) Defects in spin models. \newline
$\bullet$ {\bf(i)}  {\em Lattices viewed as graphs: } Fig. \ref{square}
shows our analysis. On the left are the result of our multi-resolution analysis. Information theory plateaus correspond to solutions on 
different scales ($\gamma$). Transitions between different solutions appear as cascades in 
the information theory measures. On the smallest scale (high $\gamma$), 
our approach recognizes the basic units of the lattices.  \newline
$\bullet$ {\bf(ii)} {\em Lattices with defects: }
In Fig. \ref{defects}, we examine a monatomic Lennard-Jones (LJ) system
(whose ground state is a triangular lattice with inter-particle spacing given by
the LJ minimum) in which defects in the form of vacancies were inserted. The system is broken 
into clusters such that the defects tend to congregate on boundaries between different 
clusters. \newline
$\bullet$ {\bf(iii)} {\em{Defects in spin systems: }} We investigated Ising systems.  
In Eq. (\ref{eq:ourPottsmodel}), we set $V_{ij}$ to be the nearest neighbor bond energies in the Ising model.
The multi-resolution analysis led to
a cascade of structures similar to that in (i) up to the largest domain wall.
Partitions into nearly perfect Ising domain walls sharply corresponded to VI {\em maxima} (and NMI minima).
This occurs as the region near the domain walls is the one which experiences the {\em largest fluctuations} in
possible assignment to the two bordering domains that it delineates. 
 These benchmark results above are reassuring: the systems ``find''  their natural structures.

\section{Applications to complex amorphous systems}
\begin{figure}
\centering
\includegraphics[width=.4\columnwidth]{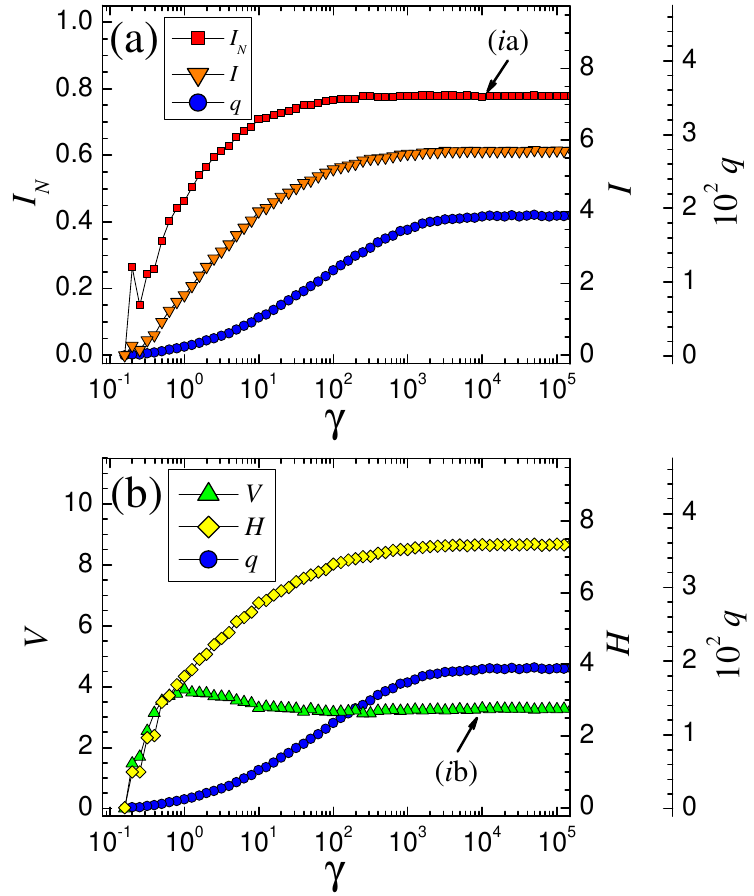}
\includegraphics[width=.4\columnwidth]{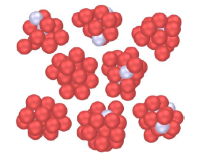}
\caption{From \cite{ronhovde}. A three dimensional KA LJ system (see text). 
Here we allow for overlapping nodes.  The information overlaps at left show a maximum NMI ($I_{N}$) and plateau for other information theory measures at (i). Right: We show the typical clusters found.}
\label{KAfig}
\end{figure}

\begin{figure}
\centering
\includegraphics[width=.45\columnwidth]{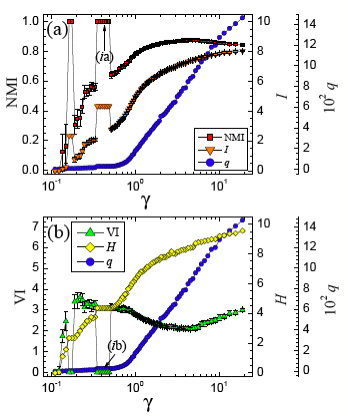}
\includegraphics[width=.75\columnwidth]{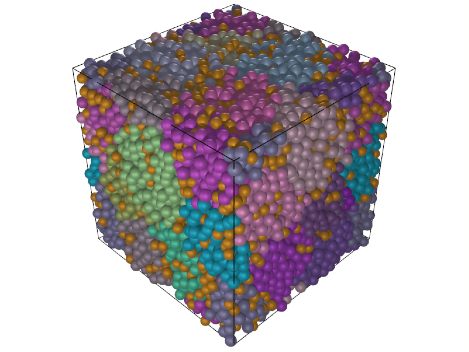}
\caption{Left:  Information theory measures applied to the Zr$_{80}$Pt$_{20}$ system (see text) of $N=10000$ atoms. The computation here were done with $V$ in Eq. \ref{eq:ourPottsmodel} replaced by the correlation functions for all different pair types (Zr-Zr, Pt-Pt, Zr-Pt) for the RMC data inferred from scattering measurements at 250K below the liquidus (1200K).  Note the sharp extrema at (i). Right: The corresponding
partition. Clusters are assigned different colors.}
\label{RMC}
\end{figure}

\begin{figure}
\centering
\includegraphics[width=.54\columnwidth]{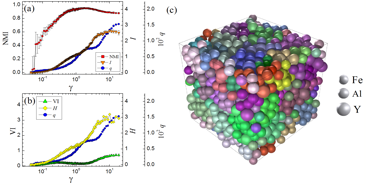}
\caption{: The result of our community detection analysis applied to
Al$_{88}$Y$_{7}$Fe$_{5}$ at a temperature of T = 300K. The panels at left (a,b) show the information theoretic overlaps between the different replicas when averaged over all replica pairs (see text). On the right (c), we highlight the spatial structures corresponding to the NMI maximum/VI minimum.}
\label{alyfe1}
\end{figure}

\begin{figure}
\centering
\includegraphics[width=.54\columnwidth]{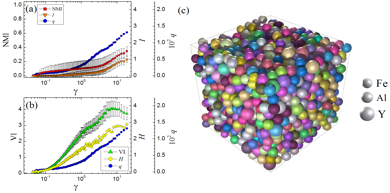}
\caption{Left:  The result of our community detection analysis as applied to Al$_{88}$Y$_{7}$Fe$_5$ 
at a temperature T = 1500K. When comparing the
information theory replica overlaps and structure with Fig. \ref{alyfe1}, 
it is evident that at higher temperatures, the system is more random. Right: corresponding structures
found at information theory extrema.}
\label{alyfe2}
\end{figure}

We studied amorphous systems via different approaches starting with that of community detection 
In particular, in analyzing complex structures we allowed in some cases for a multiple membership of a node in different communities by
further replicating each node so as to enable overlapping partitions. We used  both interaction
energies  as well as measured pair correlations as  weights  in our analysis. 
We investigated three different cases:
{\bf(i)} Static and dynamic structures in the Kob-Andersen (KA) LJ glass, {\bf(ii)} Static atomic structures
from Reverse Monte Carlo directly applied to experimental measurements of   Zr$_{80}$Pt$_{20}$, and {\bf{(iii)}} Dynamic structures in a new model amorphous system that emulates experimental results on Al$_{88}$Y$_{7}$Fe$_{5}$. 
The precise forms of the potentials in {\bf (i)} and {\bf (iii)} are of little importance in the broader context of the method. The goal is to analyze amorphous systems and find their natural sub-structures rather than where each atom happens to be  which constitutes far too detailed and hence useless information.
At low temperatures, we found  larger and more pronounced compact structures than those at higher temperature.  \newline
$\bullet$ {\bf{(i)}} We studied the glass-forming KA $80$:$20$ binary liquid 
\cite{ref:kobandersenOne,ref:valdesMixing}
by Molecular Dynamics  
(MD) \cite{ref:imd}  
to simulate an $N=2000$ atom system.  The system is initialized at a temperature $T=5$ 
(in the units of  \cite{ref:kobandersenOne,ref:valdesMixing}) and evolved 
for a time that is long compared to the caging time. We save $p$ high temperature configurations separated by time 
intervals $\Delta t$ of the order of the caging time \cite{ref:kobandersenOne}.
The system is then rapidly quenched to $T=0.01$-  well below the glass 
transition temperature of the KA-LJ system.
The system is consequently evolved  at  this lower temperature and again we save 
$p$ configurations separated by the original time interval $\Delta t$.  Each of the $p$ copies of the system
constitutes a replica within the high/low temperature problem. We then employed these replicas 
in our multi-resolution analysis. The results for the low temperature system are shown in Fig. \ref{KAfig}. Stable, relatively compact, clusters
correspond to the NMI maxima. Additional large scale domains in the low temperature
system at extrema correspond to a local NMI minimum (coincident with a VI maximum) for small 
$\gamma \simeq 0.2-0.3$. 
\newline 
$\bullet$ {\bf{(ii)}} 
Atomic configurations that are consistent with the experimentally determined scattering data for liquid Zr$_{80}$Pt$_{20}$ \cite{nakamura,saida,sordelet,wang} were generated using conventional Reverse Monte Carlo methods (RMC)  \cite{mcgreevy,keen,kim}. It is notable that RMC analysis of experimental data, unlike MD simulations,
is not limited by requisite true long equilibration times in MD simulations.
A static non-overlapping community detection partitioning with the experimentally 
measured partial pair correlations 
substituting for $V$ in Eq. (\ref{eq:ourPottsmodel}) led to extremely large clusters, at a temperature
below the liquidus (the maximum temperature at which crystals coexist with the melt).  The results
of the analysis for stable clusters are shown in Fig. \ref{RMC}.
\newline
$\bullet$ {\bf{(iii)}} Our potential energy functions for Al$_{88}$Y$_{7}$Fe$_{5}$ were computed using the techniques of \cite{ref:mihalkovicEOPP} employing ab initio results using the Vienna Ab-initio Simulation Package (VASP) \cite{vasp,vasp1,vasp2}.  The calculated structure factors were compared to experimental data \cite{sahu}.
The potentials are of the form
$V(r) = [\left(\frac{a_0}{r}\right)^{a_1} 
      + \frac{a_2}{r^{a_5} }
        \cos\left( a_3 r + a_4 \right)]$, with the parameters
 $\{a_i\}$ depending on the specific types of atom pairs (i.e., Al-Al, Al-Y, ...) (see \cite{ronhovde} and, in particular, table 1 therein for tabulated values of $\{a_{i}\}$ for the potentials; further details will be provided in \cite{effpotentials}). In Figs. (\ref{alyfe1},\ref{alyfe2}), we provide the structures of Al$_{88}$Y$_{7}$Fe$_{5}$ that our method finds at the two temperatures of T=300 K and T=1500 K.  We found that for a fixed configuration, the community detection Hamiltonian of Eq.(\ref{eq:ourPottsmodel}) with random edge weights generally exhibits a finite temperature spin-glass type transition \cite{dandan} (whereas for regular lattices Eq.(\ref{eq:ourPottsmodel}) is the standard Potts model which exhibits
 critical or first order transitions (e.g., a critical transition for $q \le 4$ on a square lattice and a first
 order transition for $q>4$)). The structures found 
 are not unique and reflect a configurational entropy 
 (different partitions may be found for a given value of the resolution 
 parameter $\gamma$ that are similar in their overall scale but different in 
 precise detail and identities of the nodes).
 These results may flesh out a facet of the 
 glass transition- as the system is supercooled-
the effective couplings become quenched. We re-iterate that, as noted above, detecting the optimally decoupled structures in more
random systems such as those that may describe the deeply supercooled  liquid constitutes a spin-glass problem. \cite{dandan}
The spin-glass transition seen
in the community detection problem for a fixed initial atomic configuration (as a function 
of temperature) enhances the change 
of partitions already evident in minimizing the Hamiltonian for the different initial atomic coordinates 
for the system at different temperatures (see, e.g., the larger structures in the low temperature system of Fig. \ref{alyfe1} vis a vis the smaller structures with much fainter information theory correlations  
at the higher temperature shown in Fig. \ref{alyfe2}). 

\section{Discussion} 

The detection of  structure in general systems is an important problem. 
We outlined a ``first principles'' network analysis method to ascertain correlations and structures
where traditional tools of analysis meet with difficulty. With the aid of this method,
we were able to detect hard to ascertain structures in complex materials such as low temperature glasses.
Aside from the canonical LJ systems used for simulations of glass former and the more specific (including experimentally driven)
efforts for the detection of metallic glass systems analyzed here, there are numerous other arenas 
which may profit from such a capability. 

We mention a few of these in passing and speculate on possible applications of our methods therein. Omnipresent memory effects and hysteresis appear in 
amorphous glass systems. \cite{memory}
These properties are of direct technological importance for data storage. ``Phase-change materials''  in the form of chalcogenide glasses appear in HD-DVD Blu-ray devices.  These devices require materials that are good glass formers, exhibit large difference in their reflectivity between the amorphous and crystalline states, and retain their amorphous state under ambient conditions. The different reflectivities of  the amorphous and crystalline states allow reading discs by measuring reflected laser light. \cite{wut,van,john,chal} Materials similar to Ge$_{8}$Sb$_{2}$Te$_{11}$ have these desired properties. Current technological progress hinges on vigorous testing of related materials in the hope of eventual improvement. Much research has also, in recent years, been done on poly-crystalline Silicon, thin films of poly-crystalline CdTe, and similar materials which can be used for photovoltaic cells. \cite{cdte}  A method such as that presented here 
may similarly examine the multi-scale structure of these materials and evolution during processing. 
The basic rudiments of the method can be generally applied to general systems having exact or effective interactions between their constituents (liquids, plasmas, etc.). 
A general classical physical system may be represented in terms of a (dynamic) 
network whose links, loops, etc. encode the measured multi-particle correlations and/or interactions.  Although we focused in this work on real space representations, analysis similar to that outlined in this work may be done for nodes that describe the system in Fourier or other spaces. 

We conclude with more speculative remarks about viable extensions of our method to electronic and other systems.
By attempting to finding an optimal partition into uncorrelated units as we have done for the 
classical systems, a multi-resolution analysis similar to devised here (and in our more elaborate companion work \cite{ronhovde}) might also be applied to quantum systems where the nodes need not conform to localized particles.  In the quantum arena, when feasible, a direct product representation in terms of
decoupled degrees of freedom (and related matrix product states \cite{mps} when these efficiently describe cluster states) 
may be quite potent. The decomposition into optimally decoupled clusters is indeed what the community
detection approach seeks to emulate in the classical systems that we have examined above.  Applications to electronic systems such as electronic glasses 
\cite{electron_glass} are natural. Other specific applications
may offer insights into the long 
sought diverging length scales (or lack thereof) in strongly correlated electronic
systems. In many of these systems, the dynamics suggests the existence of a zero temperature quantum critical point \cite{subir,my_review} yet, generally, there are no clear experimental indications of a diverging length scale. Numerous works, e.g., \cite{science_qcp,my_nature}, suggest that a quantum critical point is also present in optimally doped high temperature superconductors. Competing orders and multiple low energy states may lead to glassy response. \cite{dielectric} As in classical glasses, divergent time scales and various non-uniform structures appear in complex electronic materials. These structures are seen by crisp scanning tunneling microscope (STM) images  and other probes. \cite{davis,jan,rmp_steve,vojta,lawler}  The character of the low temperature phases remains a mystery.  Generally, a multi-resolution 
approach similar to the one outlined in this work may also be of use for theoretically analyzing general non-uniform systems where there are 
no obvious natural building blocks to consider in performing real space renormalization group (RG) type calculations and constructing coarse grained effective theories.  

\section{Acknowledgments}
We are indebted to M. Widom and M. Mihalkovi\v{c} for help with effective atomic potentials and to ongoing work \cite{effpotentials}
on the construction of these for several metallic glass formers. ZN
also wishes to thank G. Tarjus and P. G. Wolynes for critical reading
and remarks, the Lorentz Center for hosting a highly inspirational workshop (summer 2008), the KITP, and to the 
CMI of WU for partial support.


\end{document}